\documentclass[12pt]{article}
\usepackage{amsmath}
\usepackage{amsthm}
\usepackage{amssymb}
\usepackage{mathrsfs}
\usepackage{authblk}
\usepackage[usenames]{color}
\usepackage[latin5]{inputenc}
\usepackage{cite}
\usepackage{pdfsync}
\usepackage{enumitem}
\setlist{parsep=0pt,itemindent=0pt}

\theoremstyle{definition}

\numberwithin{equation}{section}
\numberwithin{thm}{section}
\numberwithin{lemma}{section}
\numberwithin{prop}{section}
\numberwithin{cor}{section}
\numberwithin{rmk}{section}
\numberwithin{defn}{section}
\numberwithin{exa}{section}

\newcommand{\gen}[1]{\partial_{#1}}
\newcommand{\pr}[1]{{\rm pr^{(#1)}}}
\newcommand{\curl}[1]{ \left\{#1\right\} }
\newcommand{\lie}{\mathfrak g}
\newcommand{\R}{\mathbb{R}}

\DeclareMathOperator{\Sl}{sl}

\DeclareMathOperator{\SL}{SL}

\DeclareMathOperator{\spn}{span}

\begin{document}
\pagenumbering{arabic}
\clearpage
\thispagestyle{empty}

\title{Lie point symmetry analysis of a second order differential equation with  singularity}

\author[1]{F.~G\"ung\"or\thanks{gungorf@itu.edu.tr}}
\author[2]{P.J. Torres\thanks{ptorres@ugr.es}}

\affil[1]{Department of Mathematics, Faculty of Science and
Letters, Istanbul Technical University, 34469 Istanbul, Turkey}
\affil[2]{Departamento de Matem\'atica Aplicada, Universidad de
Granada, 18071 Granada, Spain}

\date{}

\maketitle



\begin{abstract}
By using Lie symmetry  methods, we identify a class of second order nonlinear ordinary
differential equations invariant under at least one dimensional
subgroup of the symmetry group of the Ermakov-Pinney equation. In
this context, nonlinear superposition rule for  second order
Kummer-Schwarz equation is rediscovered. Invariance under
one-dimensional symmetry group is also used to obtain first
integrals (Ermakov-Lewis invariants). Our motivation is a type
of equations with singular term that arises in many applications, in particular in the
study of general NLS (nonlinear Schr\"odinger) equations.
\end{abstract}


\section{Introduction}

The primary motivation of this paper is to analyze the Lie point symmetries of the second order differential equation with cubic singularity
\begin{equation}\label{ermakov-1}
  \ddot{x}(t)+p(t)x(t)=q x(t)^{-3}+g(t)x(t)^m,
\end{equation}
where $q$, $m$ are constants and $p(t)$ and $g(t)$ are  arbitrary
functions of $t$. As we will see in Section \ref{sec-appl}, such
equation arises in a variety of applications coming from Classical
and Quantum Mechanics and Geometry.

In the absence of the singular term ($q=0$), the study of Lie symmetries has been performed in previous works (see \cite{Belmonte-BeitiaPerez-GarciaVekslerchikTorres2007,Belmonte-BeitiaPerez-GarciaVekslerchikTorres2008,MellinMahomedLeach1994}). On the other hand, when $g(t)\equiv 0$, we have the familiar Ermakov-Pinney (EP) equation
\begin{equation}\label{ermakov-0}
  \ddot{x}(t)+p(t)x(t)=qx(t)^{-3}.
\end{equation}
EP equation is known to be invariant under the special linear group $\SL(2,\mathbb{R})$ and admit a general solution formula (a nonlinear superposition) in terms of linearly independent solutions of the  time-dependent oscillator equation
\begin{equation}\label{oscillator}
  \ddot{x}+p(t)x=0.
\end{equation}
In a very short paper \cite{Pinney1950}, Pinney presented its general solution  as
\begin{equation}\label{pinney}
  x(t)=(Au^2+2Buv+Cv^2)^{1/2},  \qquad (AC-B^2)W^2=q
\end{equation}
where $u$ and $v$ are two independent solutions of \eqref{oscillator} and $W=u\dot{v}-v\dot{u}$ is their Wronskian (constant).
There is a vast literature devoted to this equation and many applications (for example, see \cite{Garcia-RipolPerez-GarciaTorres1999, AiChouWei2001} and \cite{LeachAndriopoulos2008} for other references).

The $\Sl(2,\mathbb{R})$ symmetry algebra  of \eqref{ermakov-0} has the basis
\begin{equation}\label{basis-sl2}
  X_1=u^2\gen t+u\dot{u}x\gen x, \quad X_2=uv\gen t+\frac{1}{2}(\dot{u}v+u\dot{v})x\gen x, \quad X_3=v^2\gen t+v\dot{v}x\gen x.
\end{equation}
They satisfy the commutation relations
\begin{equation}\label{comm}
  [X_1, X_2]=WX_1,  \quad [X_1,X_3]=2WX_2, \quad [X_2,X_3]=WX_3.
\end{equation}
$W$ can be normalized to unity by scaling the elements of the algebra. On introducing the point transformation
\begin{equation}\label{canon-1}
  \tau=\int \frac{dt}{u^2}, \qquad \xi=\frac{x}{u}, \qquad v=W\tau u
\end{equation}
we obtain, up to scaling, the canonical form of \eqref{basis-sl2}
\begin{equation}\label{comm-cano}
  \tilde{X}_1=\gen \tau,  \quad \tilde{X}_2=\tau\gen \tau+\frac{1}{2}\xi\gen \xi,  \quad \tilde{X}_3=\tau^2\gen \tau+\tau \xi\gen \xi
\end{equation}
satisfying the same commutation relations in \eqref{comm} with $W=1$. The canonical algebra \eqref{comm-cano} is one of four inequivalent representatives of Lie's classification of the algebra $\Sl(2,\mathbb{R})$ on the real plane.
The $\SL(2,\mathbb{R})$ group action on the plane $(\tau,\xi)$ is
\begin{equation}\label{SL2-act}
 (\tau,\xi)\to \left(\frac{a\tau+b}{c \tau+d},  (c \tau+d)^{-1}\xi\right), \quad  ad-bc=1.
\end{equation}
Action on $\tau$ induces a M\"obius transformation.
Transformation \eqref{canon-1} takes Eq. \eqref{ermakov-0} to $\xi''(\tau)=q\xi(\tau)^{-3}$ which is invariant under \eqref{comm-cano}.

We shall briefly review the particular case $p=q=1$, $u(t)=\cos t$, $v(t)=\sin t$ ($W=1$) which frequently arises in applications. In this case, \eqref{ermakov-0} is invariant under the $\Sl(2,\mathbb{R})$  symmetry algebra
\begin{equation}\label{basis-1}
  \begin{split}
      & X_1=\cos^2 t\gen t-\sin t\cos t x\gen x, \quad X_2=\sin t \cos t\gen t+\frac{1}{2}(\cos^2 t-\sin^2 t)x\gen x, \\
       & X_3=\sin^2 t\gen t+\sin t\cos t x\gen x.
   \end{split}
\end{equation}
From \eqref{canon-1} we find that under the transformation $\tau=\tan t$, $\xi=x\sec t$, the algebra \eqref{basis-1} is transformed to the equivalent representation \eqref{comm-cano}.
Moreover, by the change of basis $X_1+X_3\to X_1$, $X_1-X_3\to X_3$,  $2X_2\to X_2$ we have another representation of \eqref{basis-1}
$$X_1=\gen t, \quad X_2=\sin 2t\gen t+\cos 2t x\gen x,  \quad X_3=\cos 2t\gen t-\sin 2t x\gen x.$$

From \eqref{canon-1} and \eqref{SL2-act} it follows that  the $\SL(2,\mathbb{R})$ action on solutions  is given by
\begin{equation}\label{SL2-action}
  \tilde{x}(\tilde{t})=[\bigl(au(\tilde{t})-cv(\tilde{t})\bigr)^2+\bigl(bu(\tilde{t})-dv(\tilde{t})\bigr)^2)]^{1/2}x(t), \quad \tilde{\tau}=\tan \tilde{t}=\frac{a \tau+b}{c \tau+d},
\end{equation}
where $$A=\left(
            \begin{array}{cc}
              a & b \\
              c & d \\
            \end{array}
          \right)\in \SL(2,\mathbb{R}),
$$
and states that $\tilde{x}(\tilde{t})$ is a solution whenever $x(t)$ is a solution. This transformation formula can be expressed as
\begin{equation}\label{action-sol}
  \tilde{x}(\tilde{t})=[Au^2+2Buv+Cv^2]^{1/2}x(t),
\end{equation}
where we have redefined
$$A=a^2+b^2,  \quad B=-(ac+bd),  \quad C=c^2+d^2$$ so that $AC-B^2=(ad-bc)^2=1$. By \eqref{action-sol}, the constant solution $x=x_0=q^{1/4}$ is transformed to the general solution
\begin{equation}\label{gen-sol}
  x(t)=x_0 [Au^2+2Buv+Cv^2]^{1/2},
\end{equation}
where $AC-B^2=1$. This solution can be reconciled  with \eqref{pinney}  by redefining the group parameters as $(A,B,C)\to q^{1/2}(A,B,C)$ so that $AC-B^2\to q$  and the factor $x_0$ has been set to unity.   The reader is referred to Section \ref{sec2} for an alternative derivation of the general solution in the general case \eqref{ermakov-0}.

In view of the previous discussion, our first  objective is to identify a class of equations \eqref{ermakov-1} invariant under a one-dimensional subgroup of the symmetry group of the EP equation. Under this condition, the equation can be reduced to an equivalent autonomous equation by means of a canonical transformation. This study is done in Section \ref{sec2}. In Section \ref{sec3},  the most general set of nonlinear differential equations invariant under the symmetry found in Section \ref{sec2}  is identified. Section \ref{sec-EL} explores a generalization of the associated Ermakov-Lewis invariant. A related power transformation is considered in Section \ref{sec4}, rediscovering in this way the nonlinear superposition rule of the Kummer-Schwarz equation. Finally, Section \ref{sec-appl} is devoted to show some practical applications of the exposed theory to concrete models of interest in Classical and Quantum Mechanics and Geometry.

\section{Symmetries of a class of equations with singular term}\label{sec2}

In this section we analyze the Lie point symmetries of Eq. \eqref{ermakov-1}. A vector field of the form
\begin{equation}\label{vf-g}
  X=\tau(t,x)\gen t+\xi(t,x)\gen x
\end{equation}
will generate a Lie point symmetry of the equation if the second prolongation
$\pr{2}X$ of the vector field $X$  is annihilated on solutions of \eqref{ermakov-1}. Forming this infinitesimal invariance condition and splitting the resulting equation we obtain a system of linear differential equations for the coefficients $\tau$ and $\xi$. Solving this system we find that $\tau(t,x)=a(t),\xi(t,x)=c(t)x$ such that
\begin{equation}\label{a-c}
q(2c-\dot{a})=0,  \qquad 2\dot{c}=\ddot{a}
\end{equation}
and moreover $a(t)$ and $c(t)$ are linked to the coefficients $p(t),g(t)$ of the equation by the remaining determining equations
\begin{equation}\label{deteqs}
  (m-1)cg+2g\dot{a}+a\dot{g}=0, \quad \ddot{c}+2p\dot{a}+a\dot{p}=0.
\end{equation}
From \eqref{a-c}, if $q=0$ then $c(t)=\frac{1}{2}\dot{a}(t)+C_0$,
where $C_0$ is an integration constant. In
\cite{Belmonte-BeitiaPerez-GarciaVekslerchikTorres2007,Belmonte-BeitiaPerez-GarciaVekslerchikTorres2008}
the case $C_0\ne 0$ ($q=0$, $m=3$) was discussed, leading to an
autonomous equation with friction, which is not integrable in
general. In our case we assume $q\ne 0$, or $q=C_0=0$, so that our
vector field turns out to be
\begin{equation}\label{vf}
  X(a)=a(t)\gen t+\frac{1}{2}\dot{a}(t)x\gen x,  \quad a\ne 0,
\end{equation}
where $a$ must satisfy  the conditions
\begin{equation}\label{3rd}
  M(a)=\dddot{a}+4p\dot{a}+2\dot{p}a=0,
\end{equation}
and
\begin{equation}\label{cond-g}
  2a\dot{g}+(m+3)\dot{a}g=0.
\end{equation}

From Eq. \eqref{cond-g} $g(t)$ is completely determined by $a(t)$ in the form
\begin{equation}\label{g-form}
  g(t)=g_0 a^{-(m+3)/2},
\end{equation}
where $g_0$ is an integration constant. We assume $m\ne -3$ for otherwise our equation is reduced to the standard Ermakov-Pinney equation. On the other hand, the third order equation \eqref{3rd} is known to have a maximal symmetry algebra of dimension 7 for third order linear equations with an $\Sl(2,\mathbb{R})$ subalgebra \cite{KrauseMichel1988, MahomedLeach1990}. The structure of the algebra is $\lie=(\Sl(2,\mathbb{R})\oplus \mathbb{R})\vartriangleright \mathfrak{a}(3)$, where $\mathfrak{a}(3)$ is a three dimensional abelian ideal.
The $\Sl(2,\mathbb{R})$ subalgebra emerges from the fact that it has a solution basis $\curl{u^2,uv,v^2}$ with $u$ and $v$ being two linearly independent solutions of the time-dependent oscillator equation \eqref{oscillator}. In other words, the general solution of \eqref{3rd} is
\begin{equation}\label{gensol-a}
  a=Au^2+2Buv+Cv^2
\end{equation}
for arbitrary coefficients $A, B, C$. On the other hand, Eq. \eqref{3rd} has the first integral ($a$ is an integrating factor)
\begin{equation}\label{first-int}
  K=\frac{1}{4}(2a\ddot{a}-\dot{a}^2)+pa^2.
\end{equation}
If we substitute $a$  from \eqref{gensol-a} into \eqref{first-int}   we find that the constants $A, B, C$ should be constrained by $K=(AC-B^2)W^2$. Moreover, the substitution $a=y^2(t)$ transforms \eqref{first-int} to the EP equation
\begin{equation}\label{ermakov}
  \ddot{y}+py=K y^{-3}.
\end{equation}
So we have reconfirmed that the general solution of \eqref{ermakov} is given by the formula $y=\sqrt{a(t)}$.

For $g$ having the form \eqref{g-form} the original equation can be written as
\begin{equation}\label{ermakov-2}
  \ddot{x}+px=qx^{-3}+g_0x^m a^{-(m+3)/2}
\end{equation}
admitting a one-dimensional subalgebra of the symmetry algebra of the usual Ermakov-Pinney (EP) equation. Using the coordinate transformation (reparametrization of $t$ and linear change of $x$)
\begin{equation}\label{canon-2}
 \tau=\int\frac{dt}{a(t)},  \qquad y=\frac{x}{\sqrt{a}}
\end{equation}
we can write the autonomous form of \eqref{ermakov-2} as
\begin{equation}\label{aut}
\frac{d^2y}{d\tau^2}+Ky=qy^{-3}+g_0y^m.
\end{equation}
Although Eq. \eqref{aut} is exactly integrable (in the sense that it is reducible to quadratures) its full integration (general solution)  requires non-elementary functions (See Section \ref{sec-EL} for its first integral).

However, it is possible to construct group-invariant solutions as particular solutions.
From the requirement of the invariant curve condition  we must have the special solutions, up to a multiplicative nonzero constant which can be absorbed into $a(t)$, $x(t)=\sqrt{a(t)}$, where $a>0$ is given by \eqref{gensol-a}. This solution actually  corresponds to the constant solution $y=1$ of \eqref{aut}
So, under the constraint
\begin{equation}\label{constr}
  K=(AC-B^2)W^2=q+g_0,
\end{equation}
the invariant  solution of Eq. \eqref{ermakov-2} will have the form
\begin{equation}\label{nlin-super}
  x(t)=[Au^2+2Buv+Cv^2]^{1/2}.
\end{equation}
In view of \eqref{constr}, at most two parameters among $A,B,C$  can be specified so that all the parameters figuring in  solution \eqref{nlin-super}  have been fixed and consequently Eq. \eqref{nlin-super}  will represent only one particular solution.

We propose a simple example to illustrate the previous arguments. For $m=1$ Eq. \eqref{ermakov-2} is reduced to the EP equation with a modified potential
$$\ddot{x}+(p-g_0 a^{-2})x=qx^{-3}.$$  The choice $p=1$, $g_0=-\beta$, $A=1+\alpha$, $B=0$, $C=1-\alpha$ gives the two-parameter equation
$$\ddot{x}+[1+\beta(1+\alpha\cos 2t)^{-2}]x=qx^{-3}, \quad q=1+\beta-\alpha^2$$
with the special periodic solution $x=(1+\alpha\cos 2t)^{1/2}$. The harmonic oscillator equation corresponding to $q=0$ is recognised as a subclass of four-parameter Ince's equation.

The structure of the symmetry algebra for a more general variant  of the EP equation \eqref{ermakov} suggests that it can be of interest to construct general second order nonlinear ordinary differential equations invariant  under at least one-dimensional subgroup of the symmetry group of the EP equation.

\section{The most general class of invariant equations}\label{sec3}

One can construct  most general  nonlinear ordinary differential equation (ODE) of any order invariant under the symmetry \eqref{vf} for which particular group-invariant solutions can be found or reduction of order can be effected. In order to construct second order invariant ODE   we need to find a set of functionally independent second order differential invariants of the vector field \eqref{vf}, where $a(t)$ lies in the span of the solution set $\curl{u^2, uv, v^2}$. Invariants are found by solving the first order PDE defined by the condition ${\pr{2}}X(I)=0$, $I=I(t,x,\dot{x},\ddot{x})$, where
\begin{equation}\label{dif-inv-cond}
  {\pr{2}}X(a)=X(a)+\frac{1}{2}(\ddot{a}x-\dot{a}\dot{x})\gen {\dot{x}}+\frac{1}{2}(\dddot{a}x-3\dot{a}\ddot{x})\gen {\ddot{x}}.
\end{equation}
We find the first order invariants as
\begin{equation}\label{dif-inv}
  I_1=\frac{x}{\sqrt{a}}, \quad I_2=\sqrt{a}\left(\dot{x}-\frac{\dot{a}}{2a}x\right)=a\dot{I}_1.
\end{equation}
A second order differential invariant can be obtained by solving the ODE
\begin{equation}
\frac{dt}{a}=\frac{d\ddot{x}}{(\dddot{a}x-3\dot{a}\ddot{x})/2}
\end{equation}
as
\begin{equation}\label{dif-inv-3rd}
  a^{3/2}\left[\ddot{x}-\frac{1}{4a^2}(2a\ddot{a}-\dot{a}^2)x\right]=\rm{const.}
\end{equation}
Using the first integral \eqref{first-int} we find the second order invariant $I_3=a^{3/2}(\ddot{x}+px)$. An alternative to the construction of $I_3$ is to form the differentiation $dI_2/dI_1$. In general,  two functionally invariant is sufficient to determine  other higher order invariants by  the process of invariant differentiation $(D^k_x I_2)/(D^k_x I_1)$.
The most general equation admitting an invariant solution of the form $x=\sqrt{a}$, $a=Au^2+2Buv+Cv^2$ becomes
$H(I_1,I_2, I_3)=0$ with $H$ an arbitrary function. A special from is
\begin{equation}\label{inv-eq}
  \ddot{x}+px=a^{-3/2}F(I_1,I_2),
\end{equation}
where $F$ is an arbitrary function of the first two invariants. For the special choice
$F=q I_1^{-3}+g_0I_1^m$ we recover Eq. \eqref{ermakov-1} with $F(1)=q+g_0$. Another choice $F=-\alpha I_2+f(I_1)$, $\alpha$ a constant, produces the invariant equation
$$\ddot{x}+\alpha a^{-1}\dot{x}+(p-\frac{\alpha}{2}a^{-2}\dot{a})x=a^{-3/2}f(a^{-1/2}x)$$
admitting the same solution  formula  \eqref{nlin-super}: $x=\sqrt{a}$ provided $(AC-B^2)W^2=f(1)$. In general, the same rule applies to Eq. \eqref{inv-eq} with the constraint $M(a)=K=(AC-B^2)W^2=F(1,0)$. Note that no choice of the invariant $I_1$ can introduce dependence on time in the coefficient of the nonlinearity $x^{-3}$.

One can of course include higher order invariants to construct higher order nonlinear ODEs. For example, a third order invariant is given by
$$I_4=a^{5/2}[\dddot{x}+\frac{d}{dt}(px)]+\frac{3}{2}\dot{a}I_3$$ and the corresponding third order equation would have the form $H(I_1, I_2, I_3, I_4)=0$. A subclass that does not depend explicitly on $a$ and $\dot{a}$ is picked out by the  invariant equation $J_2=F(J_1)$, where $F$ is an arbitrary function and $J_1$, $J_2$ are the invariants defined in terms of $I_i$ ($i=1,2,3,4$) by
\begin{equation}\label{3rd-invariant}
  J_1=I_1^3I_3=x^3(\ddot{x}+px), \quad J_2=I_1^4(I_1I_4+3I_2I_3)=x^4[x(\dddot{x}+\dot{p}x)+\dot{x}(3\ddot{x}+4px)].
\end{equation}
The subclass $J_2=F(J_1)$ and the EP equation with basis \eqref{basis-sl2} enjoy the same $\Sl(2,\mathbb{R})$ symmetry structure. $x=\sqrt{a}$ will be a solution to this class only when  $F(K)=0$, $K=(AC-B^2)W^2$. Contrary to the second order case, this  solution provides only a two-parameter  particular solution since the equation is of third order. On the other hand, a two dimensional solvable subalgebra of the nonsolvable  $\Sl(2,\mathbb{R})$ algebra can be used to reduce its integration to a pair of quadratures and the solution of  a Riccati equation.

The change of variable  $y=x^2$ simplifies $J_1, J_2$ into a more familiar form
$$ \tilde{J}_1=\frac{1}{4}(2y\ddot{y}-\dot{y}^2)+py^2, \qquad \tilde{J}_2=\frac{y^2}{2}M(y)=\frac{y^2}{2}(\dddot{y}+4p\dot{y}+2\dot{p}y).$$
The invariant equation $\tilde{J}_2=0$ implies $M(y)=0$ (See Eq. \eqref{3rd}). We already know from Section \ref{sec2} that it is invariant under  a seven-dimensional symmetry algebra (maximal symmetry for third order linear equations), so it follows that  equation $x(\dddot{x}+\dot{p}x)+\dot{x}(3\ddot{x}+4px)=0$ ($J_2=0$) is linearizable and possesses the general solution
$$x=\sqrt{y}=\sqrt{Au^2+Buv+Cv^2}.$$

In the special case $p=0$ ($a\in\spn\curl{1,t,t^2}$) we get the invariant equation
$J_2=F(J_1)$, where $J_1=2y\ddot{y}-\dot{y}^2$, $J_2=y^2\dddot{y}$.

In terms of the invariants $r=\dot{y}$ and $s=J_1=2y\ddot{y}-\dot{y}^2$ of the solvable subalgebra $\curl{\gen t, t\gen t+y\gen y}$  it is reduced to the Riccati equation
$$\frac{dr}{ds}=\frac{s+r^2}{4F(s)}.$$  Once this equation is solved for $r=R(s)$ and inverted as $s=S(r)$ we can integrate it by two quadratures  using the fact that $s=S(r)$ is invariant under the above two-dimensional subalgebra.

We note that by the change of variable  $y=x^{-2}$ the invariants $J_1=x^3\ddot{x}$, $J_2=x^4[x\dddot{x}+3\dot{x}\ddot{x}]$  take the form
$$\tilde{J}_1=-y^{-6}[y^2\dddot{y}-6y\dot{y}\ddot{y}+6\dot{y}^2], \qquad \tilde{J}_2=-\frac{1}{2}y^{-2}[\frac{\ddot{y}}{y}-\frac{3}{2}\left(\frac{\dot{y}}{y}\right)^2].$$
The corresponding third order invariant equation is
\begin{equation}\label{olver-sl2-2-eq}
  y^2\dddot{y}-6y\dot{y}\ddot{y}+6\dot{y}^2=y^6F(s), \quad s=y^{-2}[\frac{\ddot{y}}{y}-\frac{3}{2}\left(\frac{\dot{y}}{y}\right)^2].
\end{equation}
Eq. \eqref{olver-sl2-2-eq} appeared in \cite{ClarksonOlver1996} as the canonical third order equation  invariant under  one of  the three inequivalent planar actions of the group $\SL(2,\mathbb{C})$. The authors of \cite{ClarksonOlver1996} showed  that the general solution can be formulated in parametric form in terms of solutions of the linear Schr\"odinger  equation $\psi''(\omega)-1/2 F(\omega)\psi(\omega)=0$, in addition to the alternative derivation based on  Lie reduction method. An  interesting fact that is worth mentioning is that  the symmetry algebra of Eq. \eqref{olver-sl2-2-eq}  spanned by
\begin{equation}\label{olver-sl2-2}
 \curl{\gen t, \; t\gen t-y\gen y, \; t^2\gen t-2ty\gen y}
\end{equation}
is connected to the first prolongation of the Lie algebra $\curl{\gen t, t\gen t, t^2\gen t}$ of the first action of $\SL(2,\mathbb{C})$ on the plane $(t,x)$, which projects  to the plane $(t,y)$, $y=\dot{x}$. We emphasize that all $\Sl(2,\mathbb{R})$ actions appearing throughout this paper are actually locally isomorphic to \eqref{olver-sl2-2}.

\section{Ermakov-Lewis invariants}\label{sec-EL}
Using the invariants $y=I_1$, $w=I_2$ we can reduce the order of \eqref{inv-eq} by one. Taking into account \eqref{first-int} we can write \eqref{inv-eq} in the form
\begin{equation}\label{reduced}
  w\frac{dw}{dy}+Ky=F(y,w).
\end{equation}
In general integration of \eqref{reduced} depends on the form of $F(y,w)$ which can be intractable. For example, $F(y,w)=\beta(y)w+\gamma(y)$ leads to an Abel equation of the second kind. However, integrable cases (for example when $F=\alpha(y)w^2$ we integrate a Bernoulli's equation) can generate a family of first integrals (invariants).
In particular, if $F$ is independent of $w$, the above equation is separable and can be integrated easily to give the first integral
\begin{equation}
  I=\frac{1}{2}[w^2+Ky^2]-\int^{y} F(r)dr=\text{const.},
\end{equation}
or
\begin{equation}
  I=\frac{1}{2}\left[\frac{1}{a}\left(a\dot{x}-\frac{1}{2}\dot{a}x\right)^2
  +(AC-B^2)W^2\frac{x^2}{a}\right]-\int^{x/\sqrt{a}} F(r)dr=\text{const.},
\end{equation}
where $a$ is in the linear span of the set $\curl{u^2,uv,v^2}$.
This can be viewed as a generalization of the Ermakov-Lewis invariant \cite{Lewis1967, Lewis1968}. For the special case $F=q I_1^{-3}+g_0I_1^m$ (that is, our main equation \eqref{ermakov-1}) the Ermakov-Lewis invariant is
$$I=\frac{1}{2}w^2+\frac{K}{2}y^2+\frac{q}{y^2}-\frac{g_0}{m+1}y^{m+1}, \quad m\ne -1$$ and
$$I=\frac{1}{2}w^2+\frac{K}{2}y^2+\frac{q}{y^2}-g_0\log y$$ for $m=-1$.

In particular, in the presence of $\Sl(2,\mathbb{R})$ symmetry (which occurs when $g_0=0$), by choosing
\begin{equation}\label{special-case}
  p=\omega^2, \quad u=\cos \omega t, \quad v=\sin \omega t, \quad W=\omega,  \quad g_0=0
\end{equation}
and
\begin{equation}
  \begin{split}
  & a=u^2+v^2=1, \quad K=\omega^2,\\
  & a=u^2-v^2=\cos 2 \omega t,  \quad K=-\omega^2,\\
  & a=2uv=\sin 2 \omega t,  \quad K=-\omega^2,
 \end{split}
\end{equation}
the corresponding first integrals (invariants)  are obtained as
\begin{equation}
  \begin{split}
    & I_1=H=\frac{1}{2}\dot{x}^2+\frac{\omega^2}{2}x^2+\frac{q}{2x^2}   \\
    &  I_2=[\frac{1}{2}\dot{x}^2-\frac{\omega^2}{2}x^2+\frac{q}{2x^2}]\cos 2\omega t+\omega x \dot{x} \sin 2\omega t,  \\
    &  I_3=[\frac{1}{2}\dot{x}^2-\frac{\omega^2}{2}x^2+\frac{q}{2x^2}]\sin 2\omega t-\omega x \dot{x} \cos 2\omega t,
   \end{split}
\end{equation}
where $H$ is the hamiltonian for
$\ddot{x}+\omega^2x=qx^{-3}.$
We note that these three first integrals are connected by the relation
$$H^2-I_2^2-I_3^2=q\omega^2.$$

In general, choosing $a(t)=u^2$ ($A=1$, $B=C=0$, $K=(AC-B^2)W^2=0$) we find
\begin{equation}\label{lewis-inv}
  I=\frac{1}{2}(u\dot{x}-\dot{u}x)^2-\int^{x/u} F(r)dr=\text{const.}
\end{equation}
This invariant  actually coincides with a special case of the one for the so-called Ermakov system \cite{Reid1982} defined by
\begin{equation}\label{ermakov-system}
  \ddot{x}+p(t)x=x^{-3}F\left(\frac{x}{u}\right), \qquad \ddot{u}+p(t)u=u^{-3}G\left(\frac{x}{u}\right),
\end{equation}
where $F$ and $G$ are two arbitrary functions of their arguments, when $G=0$. This special system  corresponds  exactly to \eqref{inv-eq} obtained by the replacement $F\to I_1^{-3}F(I_1)$ for $a=u^2.$

The system \eqref{ermakov-system} has the Ermakov-Ray-Reid invariant
\begin{equation}\label{ERR-invariant}
  I=\frac{1}{2}(u\dot{x}-\dot{u}x)^2-\int^{x/u} [r^{-3} F(r)-r G(r)]dr,
\end{equation}
which was originally derived by eliminating $p(t)$ between the equations
of the system \eqref{ermakov-system}
and integrating by means of the integrating factor $u\dot{x}-\dot{u}x$.


On the other hand, the system
\begin{equation}\label{ermakov-system-gen}
  \ddot{x}+p(t)x=x^{-3}F\left(\frac{x}{y}\right), \qquad \ddot{y}+p(t)y=y^{-3}G\left(\frac{x}{y}\right),
\end{equation}
preserves the $\Sl(2,\mathbb{R})$ symmetry generated by the vector fields
$$X_i=a\gen t+\frac{\dot{a}}{2}(x\gen x+y\gen y),  \quad i=1,2,3,$$ where $a=u^2,uv,v^2$ for $i=1,2,3$, respectively. The three time-dependent (functionally independent) invariants of the system \eqref{ermakov-system-gen} for $F=G=q$ (constant) are given by
\begin{equation}\label{gen-inv}
  I_i=\frac{1}{2}\left[a(\dot{x}^2+\dot{y}^2)+\frac{1}{a}\left(\frac{\dot{a}^2}{4}+K\right)(x^2+y^2)-\dot{a}(x\dot{x}+y\dot{y})
  +qa\left(\frac{1}{x^2}+\frac{1}{y^2}\right)\right].
\end{equation}

Let us mention that  the canonical coordinates \eqref{canon-2}
straightening out  the vector field \eqref{vf} to $\tilde{X}=\gen \tau$ converts the invariant equation \eqref{inv-eq}  into the autonomous form
$$\frac{d^2y}{d\tau^2}+(AC-B^2)W^2y=F\left(y,\frac{dy}{d\tau}\right).$$ In terms of $\gen \tau$ invariants $y$ and $w=dy/d\tau$ we recover Eq. \eqref{reduced} (reduction of order by one).

\section{Equivalent equations and second order Kummer-Schwarz equation}\label{sec4}
We wish to relate equations invariant under the vector fields \eqref{vf}  to other equations with the same symmetry by some power transformations of $x$.  To this end we consider vector fields
\begin{equation}\label{vf-2}
  X(a)=a(t)\gen t+k \dot{a}(t) z \gen z, \quad k\ne 0,
\end{equation}
where $a$ is assumed  as in Section \ref{sec3} and $k$ is a real constant. Vector field \eqref{vf} is equivalent to \eqref{vf-2} under the transformation $z= x^{2k}$. Second order differential invariants of \eqref{vf-2} are found as
\begin{equation}\label{dif-inv-new}
  J_1=a^{-k}z, \quad J_2=a^{-k}(a\dot{z}-k\dot{a}z),  \quad J_3=a^{2-k}(\ddot{z}+2kpz)-(2k-1)[\dot{a}J_2+\frac{k}{2}\dot{a}^2J_1].
\end{equation}
Note that in view of  \eqref{3rd} we have
$$\pr{2}X(J_3)=ka^{2-k}zM(a)=0.$$
The corresponding invariant equation will have the form $H(J_1,J_2,J_3)=0$. Solving for $J_3$ gives the class
\begin{equation}\label{inv-eq-new}
  a^{2-k}(\ddot{z}+2kpz)-(2k-1)[\dot{a}J_2+\frac{k}{2}\dot{a}^2J_1]=F(J_1,J_2),
\end{equation}
where $F$ is an arbitrary function. Thus, specifying two of the free parameters involved in $a$, say, $A, C$ (then $B$ is fixed by the following condition) within the  family of invariant equations  \eqref{inv-eq-new} we find the following formula for their particular (invariant) solutions
\begin{equation}\label{superposition}
  z(t)=a(t)^{k}=(Au^2+2Buv+Cv^2)^{k}, \quad 2k(AC-B^2)W^2=F(1,0).
\end{equation}

Formula \eqref{superposition} can give rise to a general superposition rule only when Eq. \eqref{inv-eq-new} is independent of the function $a$ and its derivative $\dot{a}$ such that the three-parameter function $a$ appears only in the symmetry transformation.
It is easy to see that this is the case for a subclass of \eqref{inv-eq-new}  when  $k=1/2$ and $F=q J_1^{-3}$. This choice leads to the Ermakov-Pinney equation \eqref{ermakov-0}. It is interesting to observe that there is  another choice of the parameter $k$ and the function $F$ which makes the functions $a$ and $\dot{a}$ disappear in \eqref{inv-eq-new}. This indeed happens when $k=-1$ and
$$F(J_1,J_2)=\frac{3}{2} \frac{J_2^2}{J_1}-2qJ_1^3,\quad q\ne 0,$$ which leads to the remarkable second order Kummer-Schwarz (2KS) equation
\begin{equation}\label{KS2}
  \ddot{z}=\frac{3}{2}\frac{\dot{z}^2}{z}+2pz-2qz^3.
\end{equation}
Eq. \eqref{KS2} arises as particular instance of the second order Gambier equation. See \cite{CarinenaGuhaLucas2013} for the details.
Apparently its $\Sl(2,\mathbb{R})$ symmetry algebra (locally isomorphic to \eqref{olver-sl2-2}) will be generated by the vector fields
\begin{equation}\label{basis-KS2}
  X_1=u^2\gen t-2u\dot{u}z\gen z,  \quad X_2=uv\gen t-(u\dot{v}+v\dot{u})z\gen z, \quad X_3=v^2\gen t-2v\dot{v}z\gen z.
\end{equation}
They satisfy the commutation relations \eqref{comm}.
We can relate the EP equation \eqref{ermakov-0} to the 2KS equation \eqref{KS2} by the power transformation $x=z^{-1/2}$, $z>0$. Apparently, the invariant solution turns into the general solution. Hence, using our approach  we have rediscovered  that Eq. \eqref{KS2} admits as its general solution the superposition rule
$$z=(A u^2+2Buv+Cv^2)^{-1}, \qquad (AC-B^2)W^2=q.$$ This fact was established in \cite{LucasSardon2012, CarinenaGuhaLucas2013} in the framework of Lie systems and quasi-Lie schemes.

2KS equation is related to a class of second order ODEs whose general solution has been obtained  as a special case by the method of  homogenous functions in \cite{Ranganath1988}. The transformation $z=w^{(n-1)/2}$, where $n\ne 1$ is a real parameter, maps Eq. \eqref{KS2} equation to
\begin{equation}\label{KS2-equiv}
\ddot{w}+\frac{4p}{1-n}w=\frac{n+3}{4}\frac{\dot{w}^2}{w}+\frac{4q}{1-n}w^n, \quad q\ne 0
\end{equation}
and preserves the $\Sl(2,\mathbb{R})$ symmetry of the 2KS equation with generators
\begin{equation}\label{basis-KS2-2}
  X_1=u^2\gen t+2ku\dot{u}w\gen w,  \quad X_2=uv\gen t+k(u\dot{v}+v\dot{u})w\gen w, \quad X_3=v^2\gen t+2kv\dot{v}w\gen w,
\end{equation}
where $k=2/(1-n)$. We observe that Eq. \eqref{KS2-equiv} can be recovered by forming  the invariant equation
$$J_3=F(J_1,J_2)=\sigma\frac{J_2^2}{J_1}+\frac{4q}{1-n}J_1^n,  \quad q\ne 0, \quad n\ne 1$$ for suitable choices of $k$ and $\sigma$. This equation can be written in the form
\begin{equation}\label{KS2-equiv-2}
  \ddot{z}+2kpz=\sigma \frac{\dot{z}^2}{z}+Q(t)\dot{z}+R(t)z+\frac{4q}{1-n}a^{k-2-nk}z^n
\end{equation}
for some functions $Q(t)$ and $R(t)$ whose exact forms will be determined below.
The coefficient of $z^n$ dictates that we must choose $k=2/(1-n)$. For this value of $k$ we have
$$Q(t)=\frac{\dot{a}}{a}\frac{(n+3-4\sigma)}{1-n},  \quad R(t)=-\frac{\dot{a}^2}{a^2}\frac{(n+3-4\sigma)}{(1-n)^2}, \quad n\ne 1.$$
We require $\sigma=(n+3)/4$ for the absence of $a$ and $\dot{a}$ in the coefficients $Q$ and $R$. Hence, Eq. \eqref{KS2-equiv-2} boils down to Eq. \eqref{KS2-equiv}.

From \eqref{superposition} we obtain the general solution of \eqref{KS2-equiv}
\begin{equation}\label{superposition-2}
 w=(A u^2+2Buv+Cv^2)^{2/(1-n)}, \qquad (AC-B^2)W^2=q.
\end{equation}
The special case $n=-3$ of \eqref{KS2-equiv} is the EP equation.
The transformation that takes \eqref{KS2-equiv} to the EP equation
$$\ddot{x}+px=qx^{-3}$$
is given by $w=x^{4/(1-n)}$.

In passing we recall that the following ODE admits the 8-dimensional $\Sl(3,\mathbb{R})$ algebra, in other words  it  is linearizable by a point transformation if and only if $n=\sigma$ or $n=1$, amounting to the fact that $g(t)$ can be replaced by zero (This necessary and sufficient condition can be justified by Lie-Tresse linearization test.):
\begin{equation}\label{linearizable}
 \ddot{w}+pw=\sigma\frac{\dot{w}^2}{w}+g(t)w^n.
\end{equation}
The transformation $w(t)=\rho(t)^{1/(1-\sigma)}$, $\sigma\ne 1$ linearizes Eq. \eqref{linearizable} to
\begin{equation}\label{linear}
  \ddot{\rho}+(1-\sigma)p\rho=(1-\sigma)g(t).
\end{equation}
For $\sigma=1$ ($g\to 0$), the transformation  and the transformed linear equation are replaced by $w=e^{\rho}$ and $\ddot{\rho}=-p$.

Finally, we note that for the choice $k=1$ ($n=-1$) in \eqref{KS2-equiv} we find the $\Sl(2,\mathbb{R})$ invariant equation
\begin{equation}\label{eq}
 \ddot{w}=\frac{1}{2}\frac{\dot{w}^2}{w}-2pw+\frac{2q}{w},
\end{equation}
which is nothing more than Eq. \eqref{first-int} with $a$ and $K$ replaced by $w$ and $q$, respectively.
From \eqref{superposition-2}
the general solution of \eqref{eq} becomes
$$w=Au^2+2Buv+Cv^2, \qquad (AC-B^2)W^2=q.$$ The transformation $w=x^2$  establishes again the connection of \eqref{eq} with the above EP equation.

\section{Applications}\label{sec-appl}

The purpose of this section is to emphasize the interest of the
Lie point symmetry analysis performed in the previous sections by
showing some examples of applications to models coming from
different scientific fields. More concretely, we will see that Eq.
$(\ref{ermakov-1})$ covers a wide range of interesting situations.

\subsection{Central force fields}

The motion of particle in a central force field is ruled by the system
\begin{equation}\label{central}
\ddot u=f(t,|u|)\frac{u}{|u|}\,,\qquad\qquad
u\in\R^N\setminus\{0\}.
\end{equation}
In a central force field every orbit is planar, hence we can assume $N=2$ without loss of generality.  Then, passing to polar coordinates $u(t)=x(t)e^{i\theta(t)}$, $x>0$ system $(\ref{central})$ is equivalent to
\begin{equation}\label{1}
\ddot x=c^2x^{-3}+f(t,x),
\end{equation}
where $c=x^2\dot\theta$ is the {\it angular
momentum} of $u$, which is a constant of motion. For $f(t,x)=g(t)x^m-p(t)x$, we recover Eq. \eqref{ermakov-1} with $q=c^2$. Once the scalar equation \eqref{1} is solved, the angular variable is found by a simple integration in $\dot\theta=cx^{-2}$. Let us write explicitly the equation
\begin{equation}\label{2}
\ddot x+p(t)x=c^2x^{-3}+g(t)x^m.
\end{equation}
We have identified a whole family of integrable equations of this type. The most general way to proceed is to fix an arbitrary function $a(t)$ and define $g(t)$ by \eqref{g-form} and
$$
p(t)=\frac{c^2+g_0}{a^2}-\frac{1}{2}[\frac{\ddot{a}}{a}-\frac{1}{2}\left(\frac{\dot{a}}{a}\right)^2].
$$
This formula for $p(t)$ is obtained from \eqref{first-int} and the restriction \eqref{constr}. For this choice of the coefficients, Eq. \eqref{2} is integrable and has $\sqrt {a(t)}$ as a particular solution and the coordinate transformation \eqref{canon-2} transforms the equation into the autonomous form \eqref{aut} with $K=c^2+g_0$, which has the energy as a conserved quantity and hence the orbits are simply the level curves of the energy.

\subsection{Modulated amplitude waves in Bose-Einstein condensates}\label{subsec4.2}

The 1D Gross-Pitaevskii equation
\begin{equation}\label{GP-1D}
iu_t = -\frac{1}{2} u_{xx} + V(x)u +
h(x)|u |^{2}u
\end{equation}
models the evolution of a quasi-one-dimensional Bose-Einstein condensate (BEC) subjected to a external magnetic trapping $V(x)$ and a particle interaction $h(x)$, which is tunable by Feshbach resonance. Assuming that both functions are $L$-periodic, a modulated amplitude wave (see \cite{Torres20143285} and the references therein) is a doubly periodic solution (in space and time), described by the ansatz
\begin{equation}\label{MAW}
u(x,t)=R(x)\exp\left(i\left[\theta(x)-\mu t\right]\right).
\end{equation}

Entering the ansatz \eqref{MAW} into \eqref{GP-1D} and taking real and ima\-gi\-nary parts, the amplitude $R(x)$ follows the second order equation
\begin{equation}\label{amp}
\ddot R(x)=\frac{c^2}{R^3}+2(V(x)-\mu) R+2h(x)R^3,
\end{equation}
where the parameter $c$ is a conserved quantity given by the
relation
\begin{equation}\label{Bose-phase}
R^2(x)\dot \theta(x)=c,
\end{equation}
in total analogy with the conservation of angular momentum of a particle under a central force field, shown in the last subsection. In fact, renaming the variables $R\to x,x\to t$, we arrive at
\begin{equation}\label{amp-x}
\ddot x+2(\mu-V(t))x=\frac{c^2}{x^3}+2h(t)x^3,
\end{equation}
which is exactly the same equation \eqref{ermakov-1} with $m=3$, $q=c^2$, $p(t)=2(\mu-V(t))$ and $g(t)=2h(t)$. Now, we can proceed as before. Of course, fixing a periodic $a(t)$ we obtain periodic coefficients. Other way to proceed is to fix $V(t)\equiv 0$, then $u=\cos \omega t,v=\sin \omega t$ with $\omega=\sqrt{2\mu}$ is a fundamental system of $\ddot x+2\mu x=0$. Taking $A=2,B=0, C=1$ in \eqref{gensol-a}, we get $a(t)=1+\cos^2 \omega t$ and $K=2\omega^2=4\mu$. Then, by \eqref{g-form} and \eqref{constr}, we can fix
$$
h(t)=(\omega^2-\frac{c^2}{2})a^{-3}.
$$
For such coefficients, \eqref{amp-x} has the periodic solution $x(t)=[1+\cos^2 \omega t]^{1/2}$ and it is reducible by the change \eqref{canon-2} to the autonomous form
$$
\ddot y+2\omega^2y=\frac{c^2}{y^3}+(2\omega^2-c^2)y^3.
$$

\subsection{The method of moments for a multi-dimensional BEC}

For the $n$-dimensional Gross-Pitaevskii equation with parabolic trap and time-dependent coefficients
\begin{equation}\label{NLS}
  iu_t=-\frac{1}{2}\Delta u+\frac{1}{2}\lambda(t)^2|x|^2 u+g(t)|u|^{2p} u,
\end{equation}
the method of moments (see \cite{GarciaTorresMontesinos2007} and the references therein) analyses the evolution of certain integral quantities with physical meaning. In particular, the second momentum (or variance)
$$
I(t)=\int_{\mathbb{R}} |x|^2 |u|^2  dx
$$
represents the width of the wave packet. It is proved in  \cite{GarciaTorresMontesinos2007} that $x(t)=\sqrt{I}$ verifies the second order equation
$$
\ddot x+\lambda(t)^2 x=q_1 x^{-3}+q_2g(t)x^m,
$$
where $q_1,q_2$ are certain conserved quantities and  $m=-(np+2)/2$.

\subsection{The 2-dimensional $L_p$ Minkowski problem}

Geometrically, the $L_p$ Minkowski problem \cite{Umanskiy2003176} asks about conditions on a function $g(x)\in C(\mathbb{S}^{n-1})$ that guarantee that it is the $L_p$ surface area measure of a convex body. When $n=2$, the problem is reduced to find a $2\pi$-periodic solution of the equation
$$
x''+x=g(t)x^{p-1}.
$$
The same equation arises in the anisotropic curve shortening problem \cite{Dohmen1996}. Observe that in this case $q=0$. Now we can proceed as in Subsection \ref{subsec4.2} to construct explicit cases of solvability. For example, for $p=-4$ the equation
$$
\ddot x+x=\frac{2(1+\cos^2 t)}{x^5}
$$
is integrable.

\bibliographystyle{unsrt}

\subsection*{Acknowledgments}
F. G\"ung\"or acknowledges the warm hospitality at the Department of
Applied Mathematics, University of Granada, Spain, where this work was started.

\end{document}